# THE STATIC QUARK-ANTIQUARK-POTENTIAL:
# A 'CLASSICAL' EXPERIMENT
# ON THE CONNECTION MACHINE CM-2[*]

K. SCHILLING, G.S. BALI

*Fachbereich Physik, Universität-Gesamthochschule Wuppertal, Gaussstr. 20*

*D-42097 Wuppertal, Germany*

This meeting produces another evidence that present parallel computers are (a) real instruments of computational physics, (b) largely in the hands of still-pioneers, (c) efficiently promoted by basic research groups with large-scale computational needs. Progress in parallel computing is carried by two types of such groups, that either follow the build-it-yourself or the early-use strategies. In this contribution, we describe, as an example to the second approach, the Wuppertal university pilot project in applied parallel computing. We report in particular about one of our key applications in theoretical particle physics on the Connection Machine CM-2: a high statistics computer experiment to determine the static quark-antiquark potential from quenched quantum chromodynamics.

## 1. Outline

In section 2 we present a brief exposé on quantum chromodynamics to non-experts, in order to illuminate our motivation to invest for considerable efforts in large-scale computing. We will also shortly mention the Wuppertal activities on parallel computing, based on the pilot installations of the Connection Machines CM-2 and CM-5. In section 3, we will illustrate that parallel computers have matured to real research tools in numerical quantum field theory. We will examplify this by presenting the state-of-the-art of Creutz' classical 1799 computer experiment, employing present hardware and software techniques in parallel computing. Section 4 contains a brief physics discussion of our results.

---

[*]Talk presented by K. Schilling at the "Workshop on Large Scale Computational Physics on Massively Parallel Computers", HLRZ, Jülich, June 14-16, 1993.

## 2. Introduction

### 2.1. *QCD — a challenge in computational physics*

Quantum chromodynamics (QCD) as the fundamental theory of strong interactions is closely related to quantum electrodynamics (QED), both being built on the principle of local gauge invariance. The gauge group of QCD is $SU(3)$ and refers to an internal charge space of three "colour" degrees of freedom.

Hadrons like the proton are composed of quarks bound by the exchange of gluons, the QCD analogs of the photons. The QCD action $S$ thus has a very appealing (since geometrical) structure, involving gluon self-couplings which make it so different from QED. The proper evaluation of QCD, nevertheless, still presents a major challenge to the theorists: their goal still is to derive the spectrum and the matrix elements of hadrons from the basic QCD action. Such *ab initio* calculations would (a) imply a clean empirical verification of the theory, which could thus be tested against a huge set of empirical low energy phenomena and (b) allow to unfold QCD effects from experimental data in future studies of the standard model of electroweak interactions.

Perturbation theory is obviously not adequate for the QCD evaluation within the low energy regime, because of the strong coupling involved; yet is has been proven to be applicable in the high energy limit of the theory, because of asymptotic freedom.

In order to go beyond perturbation theory, we remember that quantum field theories can be, through their path integral formulation, cast into a statistical mechanics problem. For this purpose, one has to continue from real to imaginary time, $t \to -it$. This Euclidean trick converts the phase factor $\exp(-iS)$ to a given "path" (field configuration) into a positive real weight, that can be readily interpreted as classical Boltzmann weight. So the technique is to estimate the (functional) path-integrals by generating a representative ensemble of field configurations $\Phi$ with Boltzmann distribution

$$P(\Phi) \simeq \exp(-S[\Phi]) \qquad (1)$$

by a stochastic process and to "measure" observables within this ensemble. After space-time discretization this amounts to Monte-Carlo summation of high dimensional integrals (dimension at least $10^5$). Discretization implies a lattice regularization – and thus a proper definition – of the original continuum quantum field theory. The lattice formulation of QCD has been introduced by Kenneth Wilson in 1974[1] and leads us directly into the arena of high performance scientific supercomputing[2] (HPSC). Normally, one chooses a hypercubic four-dimensional lattice. This lends itself easily to parallel algorithms, at least when the updating within the Markov chain is carried out locally. For more details about our actual implementation on the Connection Machine CM-2, see Ref. 3.

## 2.2. *The Wuppertal parallel computing setup*

Hopefully you can appreciate by now, that the challenge of "solving" QCD provides sufficient motivation to get involved in HPSC. Indeed, during the past years lattice gauge theory (LGT) has been an important motor for a variety of pioneering activities in parallel computing all over the world. In the University of Wuppertal the LGT group triggered four research groups from theoretical and experimental physics, electrical engineering and applied computer science into a joint initiative to launch an interdiscplinary computer laboratory for massively parallel computing in science and engineering.

In October 1990, after two years of struggling with the funding agencies, the first Connection Machine in a German university (8K CM-2 with 256 MByte memory, and a 10 GByte DATAVAULT funded by the Deutsche Forschungsgemeinschaft) was installed in Wuppertal and given into the hands of our students. Our operational style is clearly

♣ research-oriented ♣ few-group ♣ semi-dedicated ♣ educational ♣

The Connection Machines are run both by theoretical physicists and computer scientists in an informal (i.e. non computer center) style. We have developed a very nice cooperation with Thinking Machines Corporation in Boston, the manufacturer of our machines. We organize a joint seminar among the university groups involved and we are open for outside users to test their implementations on our machines.

We would like to stress, that the availability of a "home" supercomputer is of tremendous importance to the research students in computational physics: they need a supercomputer to "touch", on which they can try out their own ideas about algorithms, without an overly formal admission scheme common to supercomputer centers. Yet they have the compute power of a CRAY YMP at their disposal. In this manner, a university can fulfill its mission to challenge and train the creativity of a young generation of researchers going into computational science, which, after all, has been declared one of the goals of the US and EC grand challenge programs in HPSC.

You recognize that we are highly commited to promote HPSC in the university environment, after its previous exodus into the supercomputer centers. This is a realistic strategy, due to the advent of low price, parallel supercomputers on the market. So today, we are facing a unique chance to boost the computational approach to scientific problems by the appropriate university training.

In our case university and government agencies agreed that our pilot project should evolve with CM hardware. In this spirit we purchased a Connection Machine CM-5 (128 vector units, 1 GByte memory, 24 GByte scalable disk array), which at the time (October 1992) was the first of its kind in Germany. This pilot machine is presently performing 800 Mflops in CMFortran on a typical QCD code like Wilson fermion propagators. We expect this performance to increase at least by a factor 1.5 within the next half year.

## 3. The Creutz experiment

We think it is fair to say that Creutz' 1979 computer experiment[4] to "measure" the static $q\bar{q}$ potential opened a new era in computational hadron physics, based on the methods outlined in section . He produced first evidence that lattice techniques provide a sensible way to tackle nonperturbative effects in strong interaction physics. His conclusions in essence have been that quarks are confined, i.e. that the static potential exhibits a linear rise at large distances, and – just as important – that lattice results seem relevant to continuum physics.

Let us quickly explain how to "measure" the potential on the lattice. We start from a closed rectangular path $C(R, T)$ with extension $R \times T$. The potential is extracted from so called Wilson loops $W(R, T)$, which are defined to be the trace of path-ordered products of link variables $U_\mu(n)$ (associated to the link emanating in direction $\mu$ from lattice site $n$) along such a path $C(R, T)$. This loop construction corresponds to the world lines of a quark-antiquark pair at rest, separated by distance **R** from each other and "travelling" over a time separation $T$. In Euclidean time, this observable will reveal the static "ground state" energy for asymptotic, i.e. for sufficiently large $T$ values:

$$W(\mathbf{R}, T) = O(\mathbf{R}) \exp\{-V(\mathbf{R})T\}. \tag{2}$$

So we analyse in the interval $T \geq T_{min}$, with some reasonable cutoff $T_{min}$ and make use of the *local mass*

$$V_{T_{min}}(\mathbf{R}) = \ln\left\{\frac{W(\mathbf{R}, T_{min})}{W(\mathbf{R}, T_{min} + 1)}\right\} \tag{3}$$

as an estimator for the potential $V(\mathbf{R})$. For more details, we refer to Refs. 5,3,6.

The crucial question in the context of lattice computations is: do we see continuum physics? Or put it differently: do we observe asymptotic scaling? The answer to this question is connected to the dependence of the gauge coupling parameter, $g(a)$, on the lattice spacing, $a$, on the hypercubic lattice. In perturbation theory, this gauge coupling parameter can be shown to vanish in the limit $a \to 0$, a feature which is called asymptotic freedom. Renormalization group theory yields in the two-loop approximation for the inverse function $a(g)$:

$$a\Lambda_L = f(g) = \exp\left(-\frac{1}{2b_0 g^2}\right) (b_0 g^2)^{-\frac{b_1}{2b_0^2}}, \tag{4}$$

which form contains an energy scale, $\Lambda_L$ of the theory, such that the left hand side to this equation is dimensionless. The first two coefficients of the weak coupling expansion of the $SU(N)$ Callan-Symanzik $\beta$-function are determined in perturbation theory to be

$$b_0 = \frac{11}{3}\frac{N}{16\pi^2}, \quad b_1 = \frac{34}{3}\left(\frac{N}{16\pi^2}\right)^2. \tag{5}$$

The number of colours in our case is $N = 3$.

| $\beta$ | $S_\square$ | $\sqrt{\sigma}a$ | $\sqrt{\sigma}/\Lambda_{\overline{MS}}$ | $\sqrt{\sigma}/\Lambda_{\overline{MS}}^E$ |
|---|---|---|---|---|
| 5.5 | 0.503196( 18) | 0.5805( 84) | 4.90( 7) | 1.88( 3) |
| 5.6 | 0.475495( 27) | 0.5092(137) | 4.81( 13) | 2.12( 6) |
| 5.7 | 0.450805( 25) | 0.4066( 67) | 4.29( 7) | 2.18( 4) |
| 5.8 | 0.432349( 21) | 0.3202( 41) | 3.78( 5) | 2.12( 3) |
| 5.9 | 0.418164( 15) | 0.2530( 54) | 3.35( 7) | 1.99( 4) |
| 6.0 | 0.406318( 5) | 0.2202( 15) | 3.26( 2) | 2.02( 1) |
| 6.2 | 0.386369( 3) | 0.1581( 14) | 2.93( 3) | 1.93( 2) |
| 6.4 | 0.369364( 2) | 0.1185( 18) | 2.75( 4) | 1.89( 3) |
| 6.8 | 0.340782( 4) | 0.0694( 39) | 2.54( 14) | 1.84( 10) |

Table. 1. The average plaquette action $S_\square$, $\sqrt{\sigma}$ in lattice units, and the ratio $\sqrt{\sigma}/\Lambda_{\overline{MS}}$, calculated by use of the perturbative approximation Eq. 4 from the bare $\beta$ and an effective coupling $\beta_E = 2/S_\square$, respectively.

Note that these formulas pertain to the so called zero-flavor (quenched) sector of the theory, which neglects fermion-antifermion loops, in the spirit of the valence approximation. In these relations, $g$ stands for the so called bare lattice coupling, as it appears in the lattice action $S$. The lattice spacing $a$ is determined by matching any dimensionful empirical quantity to its respective lattice result, at a chosen $g$. With $a$ determined in this way, Eq. 4 can be used to estimate $\Lambda_L$ with different values of $g$. Asymptotic scaling is established, once the scale parameter $\Lambda_L$ is verified not to depend on $g$ any more. This is expected to occur at "sufficiently" small values of $g$. A priori, it is not clear, however, whether and where this regime can be reached in practical simulations, as we require, for the present purposes, a minimum physical lattice extent (about 1 fm) to avoid large finite size effects in the calculation, which amounts to an increasing number of lattice sites as $a \to 0$.

## 4. Physics discussion

We are interested in the static $q\bar{q}$ potential, which is expected to be dominated by a Coulomb-like term at short distances with an $R$ dependent coupling

$$V(R) \simeq V_0 - \alpha_V(aR)/R \quad \text{for } R \text{ small} \qquad (6)$$

and by a linearly rising term at large distances

$$V(R) \simeq V_0 + KR \quad \text{for } R \text{ large.} \qquad (7)$$

The coefficient $K$ is related to the empirical string tension $\sigma$ by $Ka^{-2} = \sigma = (440 MeV)^2$. A " measurement" of $K$ on the lattice thus provides the scale: the lattice spacing $a(g) = \sqrt{\sigma/K(g)}$ to a given value of the bare lattice coupling $g$. On the other hand, one might be tempted to extract from the short-distance behaviour of the potential the strength parameter

$$\alpha_V(aR) = g_V^2(aR)/(4\pi). \qquad (8)$$

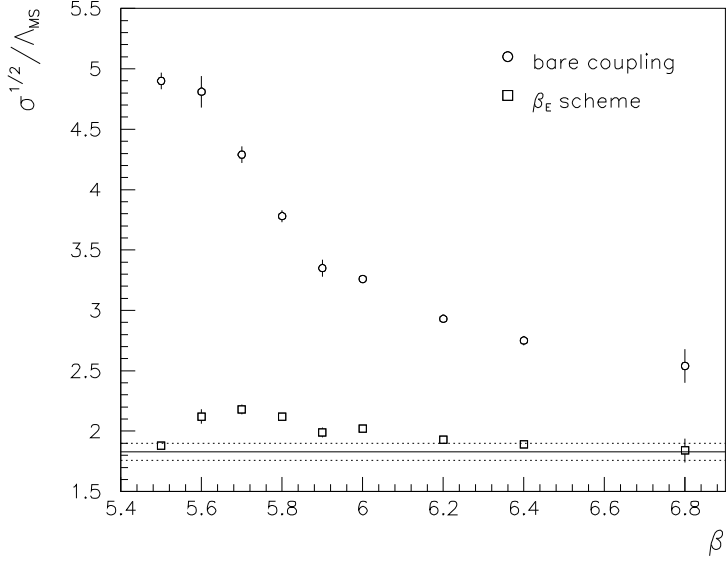

Fig. 1. The ratio $C = \sqrt{\sigma}/\Lambda_{\overline{MS}}$, in the two-loop weak coupling approximation Eq. 4, calculated by use of the bare coupling $g^2 = 6/\beta$ and the $\beta_E$-scheme $g_E^2 = 3S_\Box$. The error band denotes the value $C = 1.83 \pm 0.07$, extracted from the running of the interquark coupling $\alpha_{q\bar{q}}$.

The renormalized coupling $g_V$ can be connected to the bare lattice coupling $g$, and is due to run towards zero in a logarithmic fashion. In order to reduce the number of free parameters in this analysis, we investigate the force $F(R) = -\partial V(R)/\partial R = -\alpha_{q\bar{q}}/R^2$ rather than the potential itself. The quantity $\alpha_{q\bar{q}}$ in the two-loop approximation has the form

$$\alpha_{q\bar{q}}(aR) = \frac{1}{4\pi}\left(b_0 \ln{(Ra\Lambda_R)^{-2}} + b_1/b_0 \ln\ln{(Ra\Lambda_R)^{-2}}\right)^{-1}, \tag{9}$$

where perturbation theory relates the scales $\Lambda_R$ and $\Lambda_L$.

Physical masses like the quantity $\xi^{-1} = \sqrt{\sigma}$ can be retrieved from the lattice as inverse correlation lengths, i.e. their measurement asks for large spatial distances (in units of $a$) on top of the requirement of sufficiently large time separations $T$. So we must ascertain the inequality

$$a \ll \xi \ll La \tag{10}$$

in order to attain continuum physics. In our simulations $La$ was chosen to be larger than 1 fm.

The string tension "measurement" should be performed at large values of $R$: $K = -\lim_{R\to\infty} F(R)$. Practically, this requires large lattice volumes and huge statistics. Note however, that in reality the extraction of $\sigma$ from the charmonium spectrum is based on a length scale of $\simeq 1$ fm[†]! In both instances, one cannot neglect $1/R$ corrections to the long distance potential, Eq. 7. For this reason, it is practically impossible to achieve complete parametrization independence of the string tension analysis, in the available range of $q\bar{q}$ separations.

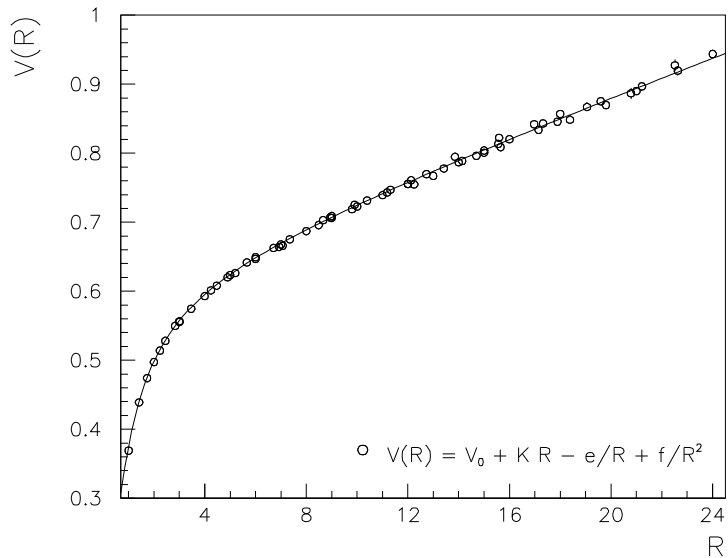

Fig. 2. The reconstructed continuum potential $V(R)$ (Eq. 15), measured at $\beta = 6.4$ as a function of $R$ with the corresponding fit curve.

In the following we will present two different approaches to a potential analysis of our data. Version I is based on the Cornell parametrization commonly used in the charmonium analysis, while version II aims at extracting continuum information from short distance data as well, where lattice artifacts have not yet passed away.

**Version I.** In the spirit of the potential analysis of quarkonia states, we start out from the Cornell ansatz for the potential[7]

$$V(R) = V_0 + KR - e/R. \qquad (11)$$

As described above, the fit range in $R$ has to be restricted. Since we are aiming at the parameter K that governs the large $R$-behaviour, we limit ourselves to the physical region $aR > 0.3$ fm, in which the experimental interquark potential has been reconstructed from the quarkonia states. We find that the last contribution to Eq. 11 can only be determined accurately for small $g$-values, $\beta = 6/g^2 > 6.0$[‡].

---

[†]Remember, that in the real world, with dynamical fermions, the confining part of the potential à la Eq. 7 will be screened by string breaking.

[‡]For $\beta < 6.0$, we have fixed $e$ to the value determined from a fit at $\beta = 6.2$, where we found $e = 0.295 \pm 0.017$

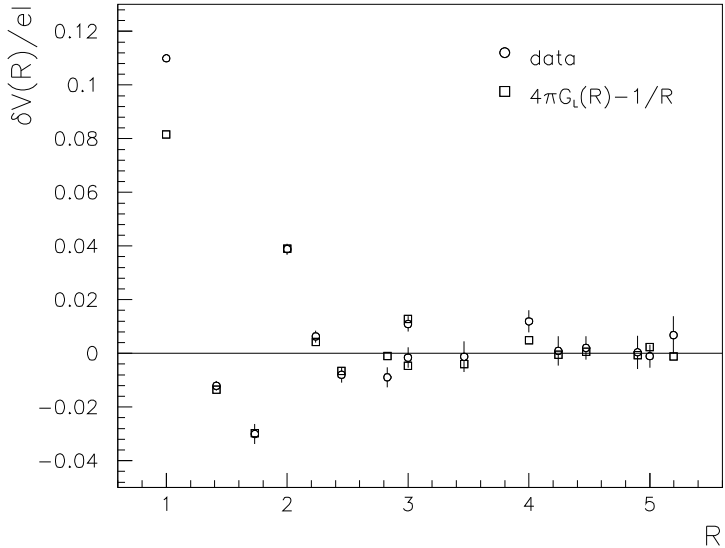

Fig. 3. Comparison between deviations of the potential at $\beta = 6.4$ from the continuum symmetry parametrization $V(R) = V_0 + KR - e/R + f/R^2$, and $4\pi G_L(\mathbf{R}) - 1/R$ (Eq. 13), which has been used to parametrize lattice artifacts.

For a scaling analysis, it is very annoying to input data from different sources, as they are subject to different systematic errors. In order to keep control on the latter, we have carried out a comprehensive study using the analysis method just described over a broad range of $\beta$: $5.5 \leq \beta \leq 6.8$. As a result, in Tab. 1 we can present the most accurate data set related to a dimensionful quantity ever determined in $SU(3)$ lattice gauge theory.

In Fig. 1 the combination $C = \sqrt{\sigma}/\Lambda_{\overline{MS}}$ is plotted vs. $\beta$, where the perturbative relation $\Lambda_{\overline{MS}} = 28.81 \Lambda_L$ has been used. $C$ has been computed via Eq. 4 (a) from the bare coupling, $28.81 C = \sqrt{K}/f(g)$ (circles) and (b) from an effective coupling, $13.88 C = \sqrt{K}/f(g_E)$ (squares) as described in Ref. 5. The error band refers to the corresponding result from the running coupling analysis presented below. We conclude, that the strong scaling violations observed in the bare lattice coupling can be considerably reduced when passing to an effective coupling scheme.

**Version II.** Here, we start from a parametrization[8] that accounts for lattice artifacts and thus enables us to make use of (more or less) the whole $R$-region:

$$V(\mathbf{R}) = V_0 + KR - e\left(\frac{1-l}{R} + l\, 4\pi G_L(\mathbf{R})\right) + \frac{f}{R^2}. \qquad (12)$$

Our analysis method has been explained in more detail in a recent paper.[5] The

lattice propagator for one-gluon exchange

$$G_L(\mathbf{R}) = \int_{-\pi}^{\pi} \frac{d^3k}{(2\pi)^3} \frac{\cos(\mathbf{kR})}{4\sum_i \sin^2(k_i/2)} \tag{13}$$

occurs inside a correction term

$$\delta V(\mathbf{R}) = el\left(4\pi G_L(\mathbf{R}) - 1/R\right), \tag{14}$$

which is supposed to render the lattice-potential $V(\mathbf{R})$ rotationally invariant:

$$V(R) = V(\mathbf{R}) + \delta V(\mathbf{R}). \tag{15}$$

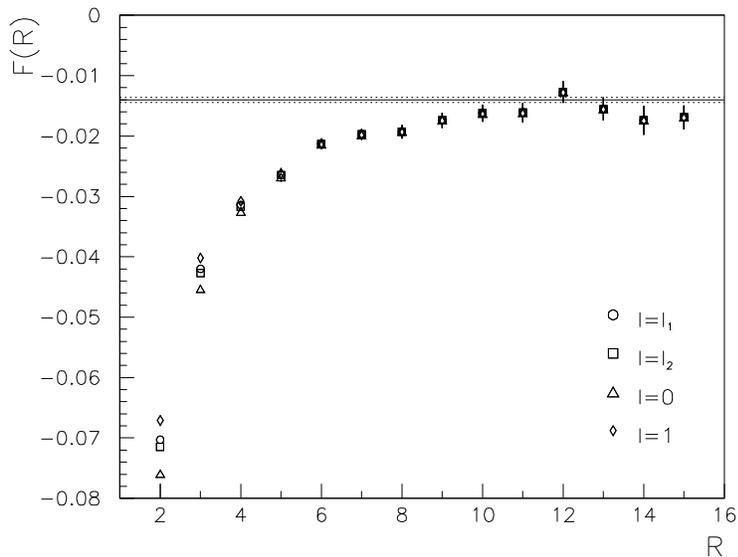

Fig. 4. The on-axis interquark force at $\beta = 6.4$ with different lattice artifact corrections. $l = 0$ denotes the case without any correction, $l = 1$ holds for a pure lattice one-gluon-exchange correction, $l_1$ denotes the "best" correction parameter obtained from the five parameter fit Eq. 12, and $l_2$ is obtained from a constrained four parameter fit with $f = 0$. The horizontal line (with dashed error) is the asymptotic value $-K$ for the force, calculated by use of the Cornell parametrization Eq. 11.

This approach is remarkably successful as can be seen in Fig. 2, where the "corrected" potential values $V(R)$ are plotted together with the corresponding fit curve. One might worry, how well the ansatz, Eq. 14, really covers the lattice artifacts. To answer this question, we compare in Fig. 3 the prediction (squares) of Eq. 14 against $\delta V(\mathbf{R})/(el)$ (circles), as computed from the data. We find, that for $R \geq \sqrt{2}$, the correction procedure works very well.[§]

---

[§] The fit incorporates five free parameters on a data set of 70 points. The fit range has been restricted to $R \geq \sqrt{3}$.

We consider next the computation of the running coupling $\alpha_{q\bar{q}}(aR)$. We start out from the corrected potential data and construct the force, in order to eliminate the $V_0$ contribution. The numerical differentiation is carried out in such a way that a pure Coulomb potential would yield a constant coupling. This is a reasonable approach to the physical situation, where the coupling is expected to vary only weakly with $R$.

We emphasize that during this stage, i.e. in the computation of the force from the corrected potential, we are independent of the previous parametrization Eq. 12, as we perform the numerical differentiation (pointwise) on the very potential data. The parametrization has only been utilized to *determine the lattice corrections*.

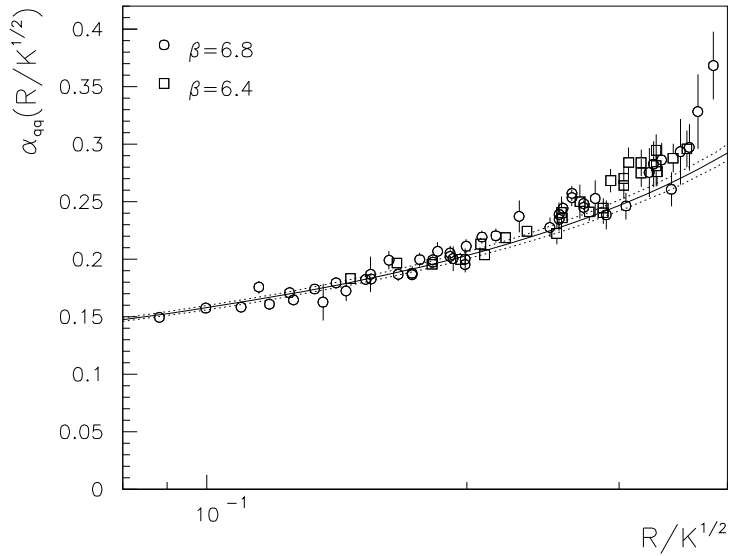

Fig. 5. The running coupling $\alpha_{q\bar{q}}$ in units of the string tension together with a one parameter fit with the two-loop formula, Eq. 9.

In Fig. 4 we display the force for the on-axis separations, where the difference between lattice (Eq. 13) and continuum propagators is most pronounced. The plot is meant to illustrate the rather weak sensitivity of the force data with respect to a reasonable change in the quantity $el$ controlling our lattice correction. We show the cases $l = 0$ (pure Coulomb, triangles), $l = 1$ (pure lattice one-gluon exchange, diamonds), and $l_1 = 0.639(35) = 0.216(11)/e_1$ (circles), the latter being our optimal value. One might worry about the impact of the $1/R^2$ term on the lattice-corrected force. For this reason, we also tried a fit with $f$ constrained to the value $f = 0$; as a result we find the value $l_2 = 0.600(58) = 0.170(16)/e_2$ (squares). The figure moreover contains the asymptotic value of the force, $-K$, as obtained from the Cornell approach (horizontal error band).

The difference between the $l_1$ and $l_2$ results will be interpreted as systematic uncertainty and is included into the error of $\alpha_{q\bar{q}}(aR)$. We note, that the $R = 2$ value of the force has not been used in the further analysis, as it involves using the potential data on $R = 1$ or $\sqrt{2}$, which might still be polluted by discretization errors.

We show the results[5] of the $\alpha_{q\bar{q}}$ analysis in Fig. 5. The curve refers to a fit to the two-loop formula Eq. 9, with $\Lambda_R = 0.572(22)\sqrt{\sigma}$. It can be seen that the short distance data on our lattices reveal the running of the coupling expected from perturbation theory. A comparison of the scale $\Lambda_{\overline{MS}}$ calculated from $\Lambda_R$ and from $\Lambda_L$ is contained in Fig. 1 and shows beautiful consistency of the short and long distance features.

## 5. Conclusion

It appears that we are finally arriving at a point where, with the available lattice techniques, we can make contact to the perturbative weak coupling regime, i.e. to continuum physics – at least in the quenched sector to QCD. This demonstrates indeed that parallel supercomputers like the Connection Machine are useful research tools in HPSC!

### Acknowledgements

The computations have been performed on the 8K Connection Machine CM-2 of the Institut für Angewandte Informatik, Wuppertal. We thank Peer Ueberholz and Randy Flesch for their enthousiastic commitment to our CM project and for constant help and advice. We are grateful to the DFG for supporting the Wuppertal CM-2 project (grant Schi 257/1-4). Our research has been funded by the EC grant # SC1*-CT91-0642.